\documentclass[conference]{IEEEtran}
\IEEEoverridecommandlockouts
\usepackage{cite}
\usepackage{amsmath,amssymb,amsfonts}
\usepackage{algorithmic}
\usepackage{graphicx}
\usepackage{textcomp}
\usepackage{comment}
\usepackage{hyperref}
\usepackage[capitalize]{cleveref}
\crefname{equation}{Eq.}{Eqs.}
\crefname{figure}{Fig.}{Figs.}
\Crefname{equation}{Eq.}{Eqs.}
\Crefname{figure}{Fig.}{Figs.}
\crefname{table}{Table}{Tables}
\Crefname{table}{Table}{Tables}
\crefname{section}{Sec.}{Secs.}
\Crefname{section}{Sec.}{Secs.}
\usepackage[braket, qm]{qcircuit}
\usepackage{physics}
\usepackage{booktabs}
\usepackage{multirow}
\usepackage{subcaption}
\usepackage{xcolor}
\def\BibTeX{{\rm B\kern-.05em{\sc i\kern-.025em b}\kern-.08em
    T\kern-.1667em\lower.7ex\hbox{E}\kern-.125emX}}
\usepackage{mathtools}
\DeclarePairedDelimiter\ceil{\lceil}{\rceil}
\DeclarePairedDelimiter\floor{\lfloor}{\rfloor}

\newcommand\change[1]{{#1}}

\begin{document}

\title{Measurement-Driven Adaptive Low-Overhead Implementation of Multi-Controlled Toffoli Gates 
}

\author{\IEEEauthorblockN{Abhoy~Kole}
\IEEEauthorblockA{\textit{Cyber-Physical Systems} \\
\textit{DFKI GmbH}\\
Bremen, Germany \\
abhoy.kole@dfki.de}
\and
\IEEEauthorblockN{Till~Schnittka}
\IEEEauthorblockA{\textit{Cyber-Physical Systems} \\
\textit{DFKI GmbH}\\
Bremen, Germany \\
till.schnittka@dfki.de}
\and
\IEEEauthorblockN{Rolf~Drechsler}
\IEEEauthorblockA{\textit{Institute of Computer Science} \\
\textit{University of Bremen / DFKI GmbH}\\
Bremen, Germany \\
drechsler@uni-bremen.de}
}




\maketitle

\begin{abstract}
The Toffoli gate is a fundamental building block for quantum arithmetic and reversible logic, yet its efficient realization remains a major challenge in both near-term and fault-tolerant quantum architectures. 
Recent advances in dynamic quantum circuit capabilities, including mid-circuit measurement and classical feedforward, provide new opportunities for reducing the resource overhead of non-Clifford operations.

In this work, we propose a set of dynamic decomposition strategies for multi-controlled Toffoli gates that exploit adaptive circuit execution and ancilla-assisted constructions. 
Our methods systematically reduce entangling-gate count, T-count, and T-depth compared with conventional static decompositions, while preserving fault-tolerance guarantees. 
Through analytical cost models and experimental evaluation, we demonstrate that relative-phase primitives and measurement-conditioned corrections enable scalable implementations with improved depth and resource efficiency.
\end{abstract}


\section{Introduction}

The Toffoli gate is a fundamental primitive in quantum computing, forming the backbone of numerous arithmetic, logical, and reversible operations. 
Its importance spans quantum algorithms~\cite{10.1145/237814.237866}, arithmetic circuits~\cite{9655478}, and the construction of universal gate sets~\cite{10.1007/3-540-10003-2_104}. 
Despite its conceptual simplicity, the efficient realization of the Toffoli gate remains a persistent challenge, particularly in the fault-tolerant regime~\cite{PhysRevA.71.022316}, where resource constraints such as qubit count, T-gate complexity, and circuit depth critically influence overall performance.

Conventional Toffoli decompositions rely heavily on non-Clifford resources, especially T gates~\cite{PhysRevA.87.042302,PhysRevA.93.022311}, which remain costly to implement and distill in contemporary quantum architectures. 
Consequently, techniques that reduce T-count, T-depth, or ancillary-qubit requirements are essential for scaling practical quantum algorithms. 
However, such optimizations often introduce trade-offs in circuit width, critical-path length, or control complexity, making efficient resource balancing an active area of research.

Recent advances in quantum hardware have enabled dynamic quantum circuit capabilities~\cite{PhysRevLett.127.100501,Martín-López2012}, including mid-circuit measurement, qubit reset, and classically conditioned operations. 
These features allow fundamentally new circuit constructions that are not accessible within strictly static models. 
By enabling adaptive control flow and conditional execution, dynamic circuits provide promising opportunities for reducing the resource overhead of non-Clifford operations, including the Toffoli gate~\cite{10546695}.

In this work, we investigate decomposition strategies that leverage dynamic circuit operations to achieve more resource-efficient Toffoli-gate implementations. 
Specifically, we propose methods that (i) minimize non-Clifford resources by reducing both T-count and T-depth in fault-tolerant settings, and (ii) exploit clean ancillas through multiple ancilla-aware optimization strategies. 
Our constructions demonstrate that incorporating mid-circuit measurement and classical feedforward can yield substantial improvements in resource efficiency, while preserving computational correctness.

Overall, this paper presents scalable Toffoli decompositions tailored to emerging hardware capabilities, providing practical pathways toward more resource-conscious quantum computation.

The remainder of this paper is organized as follows. 
Section~II reviews the relevant background, \change{Section~III} introduces the proposed methods, Section~IV presents the experimental results, and Section~V concludes the paper.

\IEEEpubidadjcol
\section{Background}
\subsection{Quantum Logic Gates}

Quantum computation is based on the manipulation of quantum bits (qubits) through a sequence of unitary transformations known as quantum logic gates. Unlike classical bits (i.e., $0$ or $1$), qubits can exist in superpositions of computational basis states (e.g., $\ket{0}$ and $\ket{1}$), enabling parallel information processing. The evolution of a closed quantum system is governed by unitary operations.
\change{From the unitary constraint, which requires that for each operation $U$ there has to be an operation $U^\dagger$ where $UU^\dagger=I$, we can derive that, unless a measurement is performed, each quantum circuit is reversible. This means that at each point in the computation, there exists a circuit that can undo this computation and return to the original input. Notably, this does not permit many operations frequently used in classical computing, such as an OR gate, which combines two input bits into a single output bit, from which the original inputs cannot be reconstructed.}

Single-qubit gates form the fundamental building blocks of quantum circuits. Common examples include the Pauli gates $X$, $Y$, and $Z$, the Hadamard gate $H$, and phase gates such as $S$ and $T$. These gates enable the preparation of superposition states (e.g., $\frac{1}{\sqrt{2}}(\ket{0}+\ket{1})$), phase manipulation, and basis transformations. When combined with multi-qubit entangling gates, such as the Controlled-NOT ($\mathrm{C}X$) gate, they provide the basis for universal quantum computation~\cite{Niel:2000}.

In practice, many quantum architectures rely on the Clifford+T gate set, consisting of Clifford gates (including $H$, $S$, and $\mathrm{C}X$) supplemented by the non-Clifford $T$ gate. While Clifford gates can be efficiently simulated classically, the inclusion of the $T$ gate enables universal quantum computation.

\begin{figure}[t]
    \centering

    \begin{subfigure}{0.175\linewidth}
        \centering
        
        \includegraphics[width=0.9\linewidth]{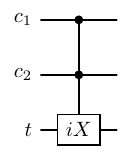}
        \vspace{0.17cm}
        \caption{}
        \label{fig:rtof1}
        
    \end{subfigure}
    \hspace{-0.4cm}
    \begin{subfigure}{0.65\linewidth}
        \centering
        \includegraphics[width=0.9\linewidth]{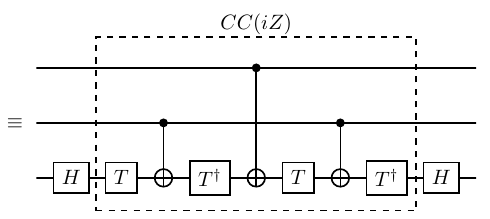}
        \caption{}
        \label{fig:rtof1-decomp}
    \end{subfigure}
    \caption{Implementations of the relative-phase Toffoli gate: (a) the $\mathrm{CC}(iX)$ operation, and (b) its Clifford+T decomposition.}
    \label{fig:rtof3}
\end{figure}

\subsection{Toffoli Gate Decomposition}

The Toffoli gate ($\mathrm{CC}X$), and its generalization to $k$ controls---the multi-controlled NOT, denoted $\mathrm{C}^{k}X$---are central multi-qubit operations in quantum computation, widely used in arithmetic circuits, reversible logic, and oracle implementations~\cite{10.1098/rsta.2023.0392,10.1007/3-540-10003-2_104,10.1145/3729229}. Despite their logical simplicity, direct physical realization of $\mathrm{C}^{k}X$ gates is challenging on most quantum hardware platforms, which natively support only single- and two-qubit interactions. Consequently, practical quantum circuits must decompose $\mathrm{C}^{k}X$ gates into sequences of elementary operations drawn from a universal gate set, such as the Clifford+T set.

The resource cost associated with such decompositions is a critical metric in circuit optimization. Common measures include the \emph{gate count}, \emph{circuit depth}, \emph{T-count}, \emph{T-depth}, and \emph{ancilla qubit overhead}. In fault-tolerant architectures, T-count and T-depth often dominate the overall cost, since non-Clifford operations require expensive ancillary-state preparation and magic state distillation~\cite{Litinski2019magicstate}. Consequently, substantial research efforts have focused on minimizing the T-count and T-depth of decomposed circuits.

Systematic optimization of $\mathrm{C}^{k}X$ synthesis---through improved decomposition techniques and ancilla-aware scheduling strategies (e.g.,~\cite{PhysRevA.87.042302,10.1007/978-3-031-38100-3_15,8758745})---directly impacts the feasibility of arithmetic-heavy algorithms and large-scale oracle-based computations.

\subsection{Dynamic Quantum Circuits}

Dynamic quantum circuits extend the conventional quantum circuit (static) model by allowing mid-circuit measurements, classical processing of measurement outcomes, and conditional quantum operations based on classical control signals. In this paradigm, the execution of a circuit may branch depending on observed outcomes, enabling adaptive decision-making within a single computational run. 

The availability of mid-circuit measurement and feedforward enables a wide range of advanced techniques, including synthesizing \change{quantum} circuits limited qubits,
active error detection and correction, teleportation-based gate implementations, adaptive phase estimation, and measurement-assisted gate synthesis~\cite{PhysRevResearch.7.023120, PhysRevLett.127.100501,10137250,doi:10.1126/science.aaw9415}. In the context of multi-controlled operations, such capabilities can be exploited to replace long coherent gate sequences with shorter adaptive procedures, thereby reducing quantum depth and accumulated noise.



In this work, dynamic circuit techniques are leveraged to optimize the implementation of multi-controlled Toffoli gates. By exploiting mid-circuit measurement and classical control, the proposed approach reduces coherent gate depth and non-Clifford resource consumption, thereby improving scalability for arithmetic-intensive and oracle-based quantum algorithms.


\section{Dynamic Clifford+T Optimization}\label{sec:dynamic-optimization}
\begin{figure}[t!]
    \centering

    \begin{subfigure}{0.17\linewidth}
        \centering
        
        \includegraphics[width=0.81\linewidth]{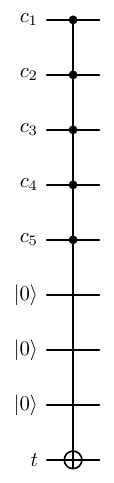}
        \vspace{0.0cm}
        \caption{}
        \label{fig:tof6}
        
    \end{subfigure}
    \hspace{-0.4cm}
    \begin{subfigure}{0.71\linewidth}
        \centering
        \includegraphics[width=0.905\linewidth]{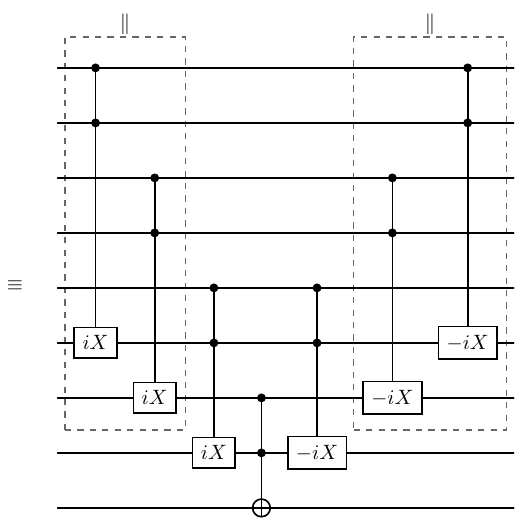}
        \caption{}
        \label{fig:tof6-decomp}
    \end{subfigure}

    \caption{Clean-ancilla decomposition of the $\mathrm{C}^{5}X$ gate: (a) the $\mathrm{C}^{5}X$ operation, and (b) its Clifford+T decomposition using $\mathrm{CC}(iX)$ gates.}
    \label{fig:tof-decomp}
\end{figure}
\subsection{Dynamic Optimization Using $\mathrm{CC}(iX)$ Gates}
Existing Clifford+T realizations of the three-input Toffoli ($\mathrm{CC}X$) gate require six $\mathrm{C}X$ gates and seven $T$ gates, with a $T$-depth of three, as reported in~\cite{6516700}. 
Selinger~\cite{PhysRevA.87.042302} proposed an alternative construction achieving a $T$-depth of one, at the cost of introducing four ancillary qubits. 
Furthermore, it has been shown that relative-phase Toffoli gates can offer significant advantages in the synthesis of larger multi-controlled Toffoli gates ($\mathrm{C}^{n}X$ for $n>2$)~\cite{PhysRevA.93.022311}.

\Cref{fig:rtof3} presents Clifford+T implementations of three-input relative-phase Toffoli gates ($\mathrm{CC}(iX)$). 
Using these primitives, a $\mathrm{C}^{n}X$ gate can be constructed with $n-2$ ancillary qubits, as illustrated in~\Cref{fig:tof-decomp} for the case $n=5$. 
The $\mathrm{CC}(iX)$ implementation shown in~\Cref{fig:rtof1-decomp} requires three $\mathrm{C}X$ gates, four $T$ gates, and has a $T$-depth of four. 
Consequently, a $\mathrm{C}^{n}X$ circuit synthesized from $n-2$ clean ancillas, $2(n-2)$ $\mathrm{CC}(iX)$ gates, and one $\mathrm{CCX}$ gate incurs the following total resource cost:
\begin{align}\label{eq:tofn-decomp}
\mathcal{C}_{\mathrm{ost}}(\mathrm{C}^{n}X, n-2) 
&= 6(n-1)~\mathrm{C}X  \nonumber\\
&\quad + (8n-9)~\text{$T$-count} \nonumber\\
&\quad + \big(8\ceil*{\tfrac{n}{2}}-5\big)~\text{$T$-depth}.
\end{align}

The reduced $T$-depth is achieved through parallel execution of $\mathrm{CC}(iX)$ gates acting on disjoint qubit subsets, enabled by appropriate scheduling, as illustrated in~\Cref{fig:tof6-decomp}.

\begin{figure}[t!]
    \centering     
        \includegraphics[width=0.89\linewidth]{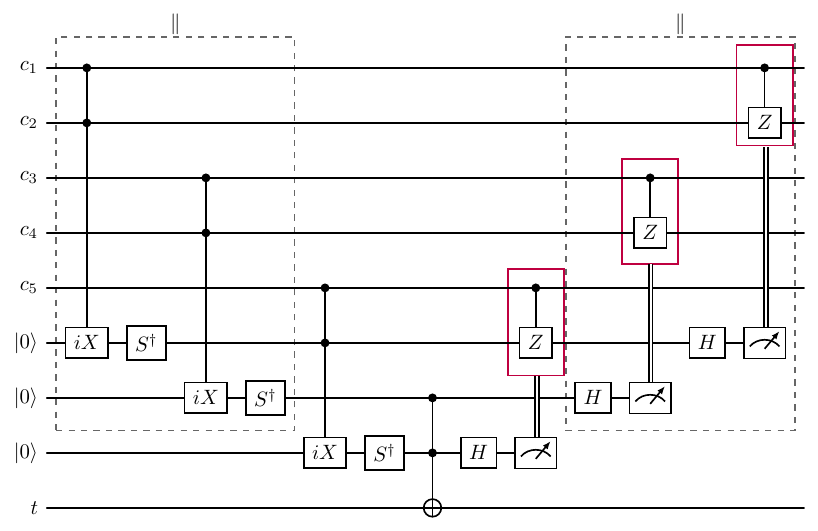}
    \caption{Clean-ancilla–based dynamic decomposition of the $\mathrm{C}^{5}X$ gate, using three $\mathrm{CC}(iX)$ operations followed by three measurement-conditioned $\mathrm{C}Z$ corrections.}
    \label{fig:tof-dynamic-decomp}
\end{figure}

Building on the dynamic and fault-tolerant $\mathrm{CCX}$ construction proposed in~\cite{PhysRevA.87.022328}, which achieves a T-count of four using a single clean ancilla, a similar strategy can be extended to the dynamic realization of $\mathrm{C}^{n}X$ gates. 
In particular, the $\mathrm{CC}(-iX)$ components (see~\Cref{fig:tof6-decomp}) appearing in the standard $\mathrm{C}^{n}X$ decomposition can be replaced by a sequence consisting of ancilla phase correction ($S^\dagger$), measurement in the Hadamard basis ($M_H$), and measurement-conditioned phase restoration ($\mathrm{C}Z$). 
This transformation enables the use of mid-circuit measurement and classical feedforward to reduce coherent non-Clifford resources.

As a result, the overall resource cost of the dynamic $\mathrm{C}^{n}X$ implementation is given by
\begin{align}\label{eq:tofn-decomp-dynamic}
\mathcal{C}_{\mathrm{cost}}(\mathrm{C}^{n}X, n-2) 
= \mathcal{C}^{\mathcal{S}}_{\mathrm{cost}} 
+ \mathcal{C}^{\mathcal{D}}_{\mathrm{cost}},
\end{align}
where the static cost is
\begin{align}\label{eq:tofn-static-cost}
\mathcal{C}^{\mathcal{S}}_{\mathrm{cost}}(\mathrm{C}^{n}X, n-2) 
&= 3n~\mathrm{C}X  \nonumber\\
&\quad + (4n-1)~\text{$T$-count} \nonumber\\
&\quad + \big(4\ceil*{\tfrac{n}{2}}-1\big)~\text{$T$-depth}.
\end{align}
and the dynamic cost is
\begin{align}\label{eq:tofn-dynamic-cost}
\mathcal{C}^{\mathcal{D}}_{\mathrm{cost}}(\mathrm{C}^{n}X, n-2) 
&= 
\begin{cases}
(n-2)~\mathrm{C}X, & M_H^{\otimes (n-2)} = 2^{n-2}-1, \\
0,  & M_H^{\otimes (n-2)} = 0,
\end{cases}
\end{align}
where $M_H^{\otimes (n-2)}$ denotes the outcome of measuring all $n-2$ qubits in the Hadamard basis.
\change{Depending on the architecture, measurement, classical control and reset might incur additional costs. However, these are not considered here.}

\Cref{fig:tof-dynamic-decomp} illustrates the resulting realization of the $\mathrm{C}^{5}X$ gate. 
Compared with the static construction in~\Cref{eq:tofn-decomp}, this dynamic approach yields reductions of 12 $T$ gates and eight units of $T$-depth (and six $\mathrm{C}X$ gates in the worst case, when all measured ancilla outcomes are 1), thereby demonstrating the substantial resource savings enabled by adaptive circuit execution.

\subsection{Dynamic Optimization Using $\mathrm{C}^{3}(iX)$ Gates}

\begin{figure}[t]
    \centering
        \includegraphics[width=0.95\linewidth]{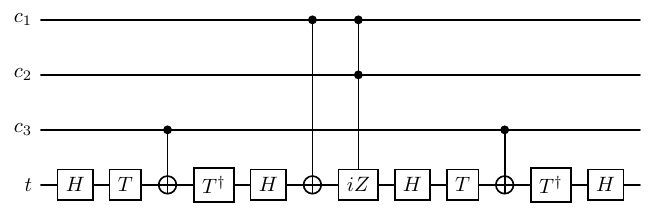}
    \caption{The Clifford+T realization of the $\mathrm{C}^{3}(iX)$ gate.}
    \label{fig:rtof4}
\end{figure}

According to~\cite{PhysRevA.93.022311}, the use of the four-input relative-phase Toffoli gate, denoted $\mathrm{C}^{3}(iX)$, enables further simplification of the Clifford+T realization of $\mathrm{C}^{n}X$ gates for $n>3$. 
As illustrated in \Cref{fig:rtof4}, the Clifford+T implementation of $\mathrm{C}^{3}(iX)$ requires twice the resources of a $\mathrm{CC}(iX)$ gate, namely six $\mathrm{C}X$ gates, eight $T$ gates, and a $T$-depth of eight.
Incorporating $\mathrm{C}^{3}(iX)$ gates into the construction of $\mathrm{C}^{n}X$ reduces the upper bound on the number of required clean ancillas from $n-2$ to $\ceil*{\tfrac{n-2}{2}}$. 

\subsubsection{Even $n$ Case}
The resulting construction consists of $n-2$ $\mathrm{C}^{3}(iX)$ gates and one $\mathrm{CC}X$ gate, yielding the following resource cost:
\begin{align}\label{eq:tofn-decomp-II}
\mathcal{C}_{\mathrm{cost}}\big(\mathrm{C}^{n}X, \tfrac{n-2}{2}\big) 
&= 6(n-1)~\mathrm{C}X  \nonumber\\
&\quad + (8n-9)~\text{$T$-count} \nonumber\\
&\quad + \big(16\floor*{\tfrac{n}{4}}+3\big)~\text{$T$-depth}.
\end{align}

\begin{figure}[t]
    \centering

    \begin{subfigure}{0.2\linewidth}
        \centering
        \includegraphics[width=0.73\linewidth]{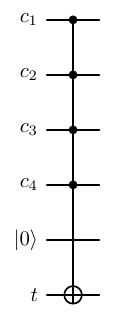}
        \caption{}
        \label{fig:tof5}
        
    \end{subfigure}
     \hspace{-0.1cm}
    \begin{subfigure}{0.351\linewidth}
        \centering
        \includegraphics[width=0.95\linewidth]{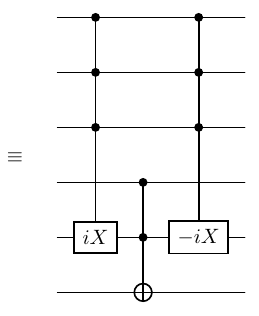}
        \caption{}
        \label{fig:tof5-decomp-2}
    \end{subfigure}

    \caption{Clean-ancilla decomposition of the $\mathrm{C}^{4}X$ gate: (a) the $\mathrm{C}^{4}X$ operation, and (b) its Clifford+T decomposition using $\mathrm{C}^{3}(iX)$ gates.}
    \label{fig:tof5-decomp}
\end{figure}

An explicit realization of this construction for $\mathrm{C}^{4}X$ is shown in \Cref{fig:tof5-decomp}, which conforms to~\cref{eq:tofn-decomp-II}.

Following the dynamic replacement strategy introduced in the previous subsection, the present construction extends this approach to the $\mathrm{C}^{3}(-iX)$ components appearing in the static decomposition (see~\Cref{fig:tof5-decomp-2}). 
These components are realized using a similar adaptive sequence based on ancilla phase correction ($S^\dagger$), measurement in the Hadamard basis ($M_H$), and measurement-conditioned phase restoration.

\begin{figure}[t]
    \centering
    \includegraphics[width=0.7\linewidth]{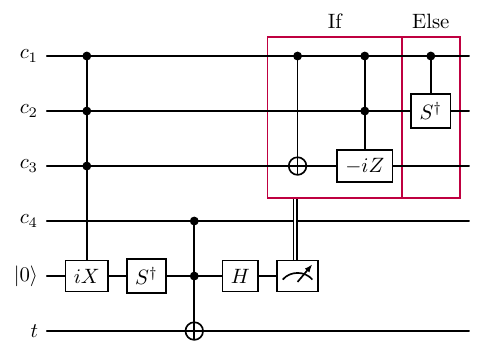}
    \caption{A dynamic implementation of the $\mathrm{C}^{4}X$ gate using one $\mathrm{C}^{3}(iX)$ gate together with measurement-conditioned operations: $\mathrm{C}X\cdot \mathrm{CC}(-iZ)$ when the ancilla outcome is 1, and $\mathrm{C}S^\dagger$ when it is 0.}
    \label{fig:tof5-dynamic-2}
\end{figure}

Depending on the measurement outcome, the required correction is implemented by either $\mathrm{C}X\cdot \mathrm{CC}(iZ)$ or $\mathrm{C}S^\dagger$, as illustrated in \Cref{fig:tof5-dynamic-2}. 
This extension preserves the fault-tolerant and resource-efficient properties of the underlying dynamic $\mathrm{C}^{n}X$ construction while enabling further reductions in coherent non-Clifford overhead.

Assuming $\mathrm{C}X\cdot \mathrm{CC}(-iZ)$ as the worst-case dynamic overhead, the adaptive implementation requires $\tfrac{n-2}{2}$ $\mathrm{C}^{3}(iX)$ gates, $\tfrac{n-2}{2}$ conditional correction blocks, and one $\mathrm{CC}X$ gate. 
This results in the following static cost:
\begin{align}\label{eq:tofn-decomp-II-static-dynamic}
\mathcal{C}^{\mathcal{S}}_{\mathrm{cost}}\big(\mathrm{C}^{n}X, \tfrac{n-2}{2}\big) 
&= 3n~\mathrm{C}X  \nonumber\\
&\quad + (4n-1)~\text{$T$-count} \nonumber\\
&\quad + \big(8\floor*{\tfrac{n}{4}}+3\big)~\text{$T$-depth}.
\end{align}

The corresponding dynamic cost is given by
\begin{align}\label{eq:tofn-decomp-II-dynamic}
\mathcal{C}^{\mathcal{D}}_{\mathrm{cost}}\big(\mathrm{C}^{n}X, \tfrac{n-2}{2}\big) 
=\qquad\qquad\qquad\qquad\qquad\qquad\quad\qquad\nonumber \\
\begin{cases}
\begin{aligned}
& (2n-4)~\mathrm{C}X  \\
& + (2n-4)~\text{$T$-count} \\
& + 4\floor*{\tfrac{n}{4}}~\text{$T$-depth},
\end{aligned}
& M_H^{\otimes \ceil*{\frac{n-2}{2}}} = 2^{\ceil*{\frac{n-2}{2}}}-1,
\\[16pt]
\begin{aligned}
& (n-2)~\mathrm{C}X  \\
& + 3\big(\tfrac{n-2}{2}\big)~\text{$T$-count} \\
& + 2\floor*{\tfrac{n}{4}}~\text{$T$-depth},
\end{aligned}
& M_H^{\otimes \ceil*{\frac{n-2}{2}}} = 0,
\end{cases}
\end{align}
where $M_H^{\otimes \ceil*{\frac{n-2}{2}}}$ denotes the joint measurement outcome of the $\ceil*{\tfrac{n-2}{2}}$ ancilla qubits in the Hadamard basis.

\subsubsection{Odd $n$ Case}

For odd values of $n$, the proposed construction employs $n-3$ $\mathrm{C}^{3}(iX)$ gates, two $\mathrm{CC}(iX)$ gates, and one $\mathrm{CC}X$ gate. 
The resulting resource cost is given by
\begin{align}\label{eq:tofn-decomp-III}
\mathcal{C}_{\mathrm{cost}}\big(\mathrm{C}^{n}X, \tfrac{n-2}{2}\big) 
&= 6(n-1)~\mathrm{CNOT}  \nonumber\\
&\quad + (8n-9)~\text{$T$-count} \nonumber\\
&\quad + (4n-1)~\text{$T$-depth}.
\end{align}

\begin{figure}[t]
    \centering
    \includegraphics[width=0.9\linewidth]{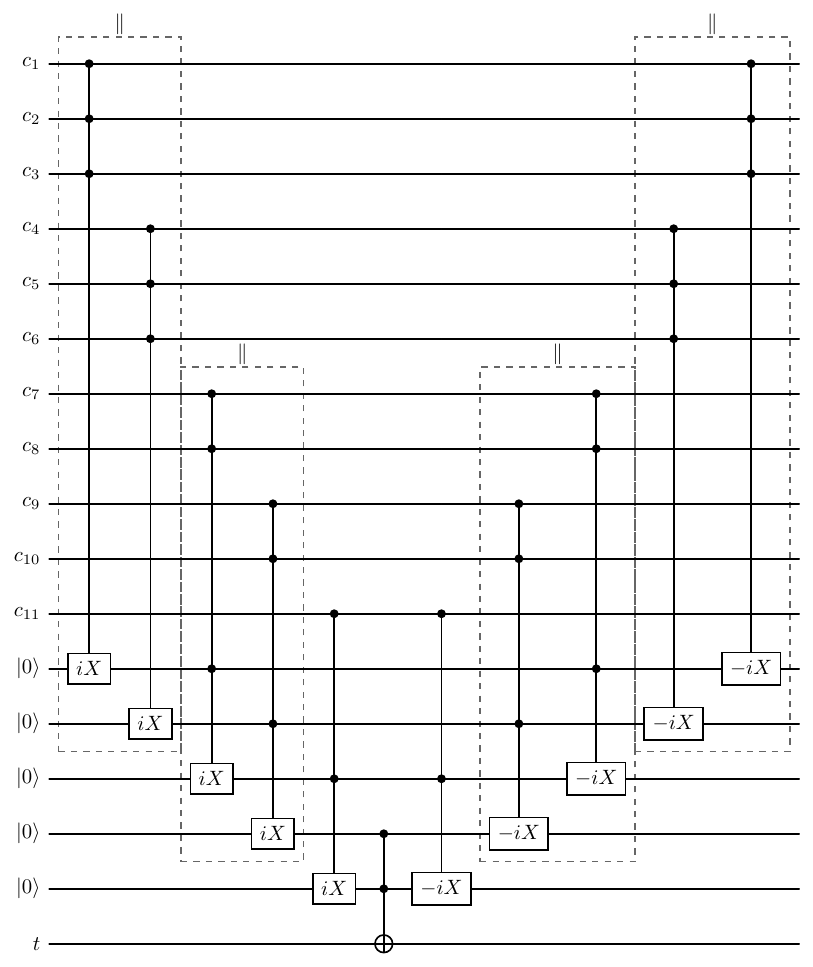}

    \caption{Clean-ancilla decomposition of the $\mathrm{C}^{11}X$ gate using $\mathrm{C}^{3}(iX)$ and $\mathrm{CC}(iX)$ gates.}
    \label{fig:tof11-decomp}
\end{figure}

An explicit realization of this construction for $\mathrm{C}^{11}X$ is shown in \Cref{fig:tof11-decomp}, which conforms to~\cref{eq:tofn-decomp-III}.

\begin{figure*}[t!]
    \centering     
        \includegraphics[width=0.9\linewidth]{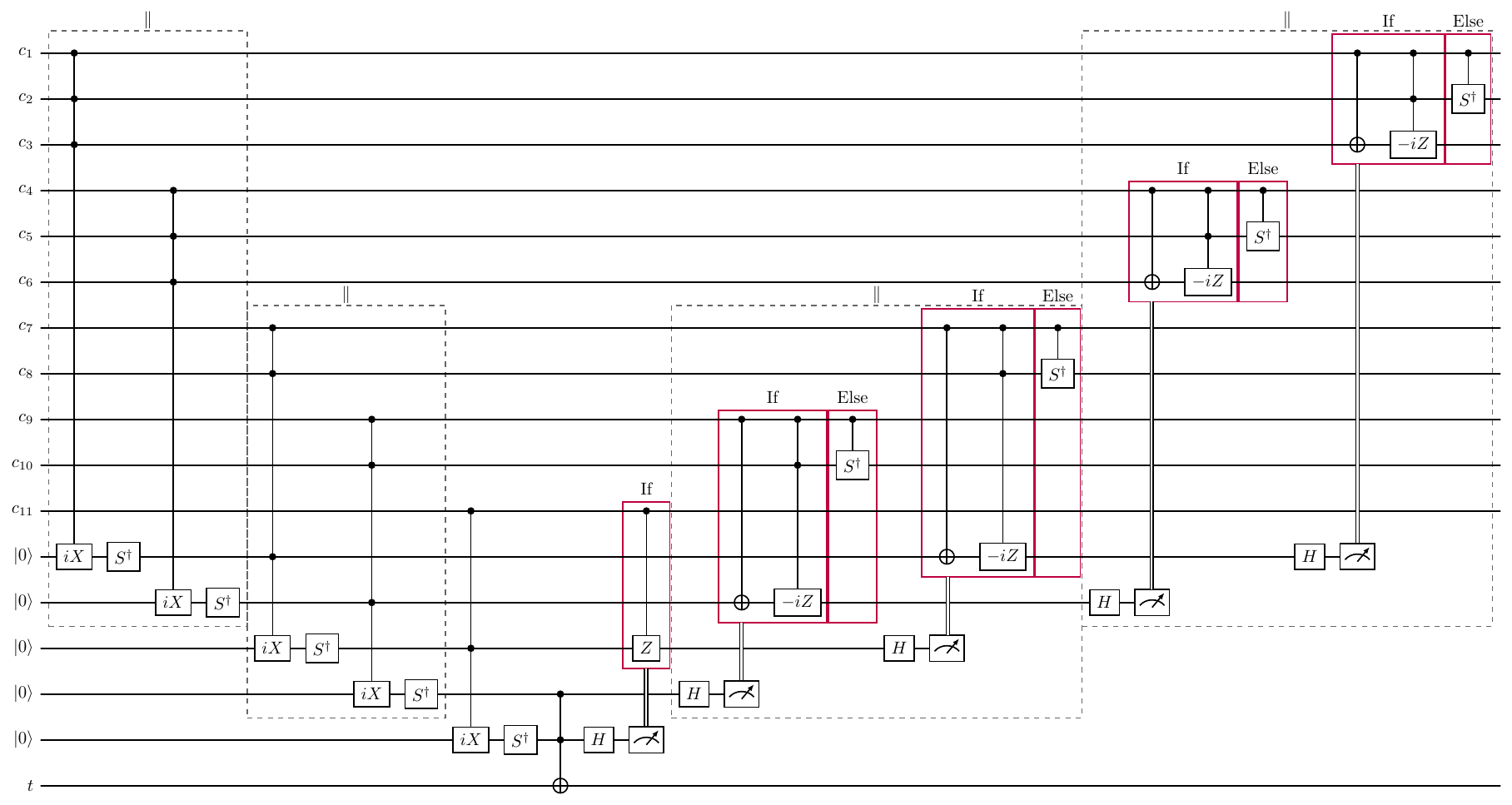}
    \caption{Clean-ancilla–based dynamic decomposition of the $\mathrm{C}^{11}X$ gate: 
the construction employs four $\mathrm{C}^{3}(iX)$ operations and one $\mathrm{CC}(iX)$ operation, 
followed by five measurement-conditioned phase corrections. 
For $\mathrm{CC}(iX)$, the correction is implemented by $\mathrm{C}Z$, 
while for $\mathrm{C}^{3}(iX)$ it is given by either $\mathrm{C}X\cdot \mathrm{CC}(-iZ)$ or $\mathrm{C}S^\dagger$, depending on the measurement outcome.
}
    \label{fig:tof11-dynamic-decomp}
\end{figure*}

Following the dynamic replacement strategy introduced earlier, this construction is extended to the $\mathrm{CC}(-iX)$ and $\mathrm{C}^{3}(-iX)$ components appearing in the static decomposition (see~\Cref{fig:tof11-decomp}). 
For $\mathrm{CC}(-iX)$, the required correction is realized by $\mathrm{C}Z$, whereas for $\mathrm{C}^{3}(-iX)$ it is implemented by either $\mathrm{C}X\cdot \mathrm{CC}(iZ)$ or $\mathrm{C}S^\dagger$, depending on the measurement outcome, as illustrated in \Cref{fig:tof11-dynamic-decomp}. 

Assuming $\mathrm{C}X\cdot \mathrm{CC}(-iZ)$ as the worst-case dynamic overhead for $\mathrm{C}^{3}(-iX)$, the adaptive implementation requires $\tfrac{n-3}{2}$ $\mathrm{C}^{3}(iX)$ gates, one $\mathrm{CC}(iX)$ gate, and a comparable number of conditional correction blocks, in addition to one $\mathrm{CC}X$ gate. 
The corresponding static cost is therefore given by
\begin{align}\label{eq:tofn-decomp-III-static-dynamic}
\mathcal{C}^{\mathcal{S}}_{\mathrm{cost}}\big(\mathrm{C}^{n}X, \tfrac{n-2}{2}\big) 
&= 3n~\mathrm{C}X  \nonumber\\
&\quad + (4n-1)~\text{$T$-count} \nonumber\\
&\quad + (2n+1)~\text{$T$-depth}.
\end{align}

The corresponding dynamic cost is given by
\begin{align}\label{eq:tofn-decomp-III-dynamic}
\mathcal{C}^{\mathcal{D}}_{\mathrm{cost}}\big(\mathrm{C}^{n}X, \tfrac{n-2}{2}\big) 
=\qquad\qquad\qquad\qquad\qquad\qquad\quad\qquad\nonumber \\
\begin{cases}
\begin{aligned}
& (2n-5)~\mathrm{C}X  \\
& + (2n-6)~\text{$T$-count} \\
& + (n-1)~\text{$T$-depth},
\end{aligned}
& M_H^{\otimes \ceil*{\frac{n-2}{2}}} = 2^{\ceil*{\frac{n-2}{2}}}-1,
\\[14pt]
\begin{aligned}
& (n-3)~\mathrm{C}X  \\
& + 3\big(\tfrac{n-3}{2}\big)~\text{$T$-count} \\
& + 2\floor*{\tfrac{n-1}{4}}~\text{$T$-depth},
\end{aligned}
& M_H^{\otimes \ceil*{\frac{n-2}{2}}} = 0,
\end{cases}
\end{align}
where $M_H^{\otimes \ceil*{\frac{n-2}{2}}}$ denotes the joint measurement outcome of the $\ceil*{\tfrac{n-2}{2}}$ ancilla qubits in the Hadamard basis.

The dynamic realization of the $\mathrm{C}^{11}X$ gate shown in~\Cref{fig:tof11-dynamic-decomp} conforms to the static and dynamic resource estimates in~\cref{eq:tofn-decomp-III-static-dynamic,eq:tofn-decomp-III-dynamic}.

\section{Experimental Results}
For the experimental evaluation, Toffoli gates with up to 16 qubits were considered. 
Each gate was decomposed \change{according to the above-described constructions} using two clean-ancilla-based approaches: 
(i) a construction employing $n-2$ clean ancilla qubits and 
(ii) a construction using $\ceil*{\tfrac{n-2}{2}}$ clean ancillas. 
The results are summarized in \Cref{tab1} and compared with state-of-the-art clean-ancilla methods reported in~\cite{Niel:2000,PhysRevA.93.022311}.

To ensure a fair comparison, the resource overhead in terms of $\mathrm{C}X$ count, T-count, and T-depth for the proposed scheme is evaluated by combining the static cost $\mathcal{C}^{\mathcal{S}}_{\mathrm{cost}}$ with the worst-case dynamic overhead, corresponding to $\mathcal{C}^{\mathcal{D}}_{\mathrm{cost}}$ for $M_H^{m}=2^{m}-1$.
\change{For the purpose of this comparison, we look at T-depth on the level of scheduled Toffoli gates and do not consider T-depth scheduling beyond this.}

The results indicate that the proposed dynamic decomposition consistently reduces the $\mathrm{C}X$ count, T-count, and T-depth compared with both static constructions. 
In particular, while the static approach based on $\ceil*{\tfrac{n-2}{2}}$ clean ancillas yields circuit costs in terms of $\mathrm{C}X$ and T-count comparable to those of the $n-2$ ancilla construction, it incurs a higher T-depth.

By contrast, the proposed dynamic scheme achieves simultaneous reductions in $\mathrm{C}X$, T-count, and T-depth under both ancilla configurations. 
Moreover, the lowest overall resource overhead is obtained when $n-2$ clean ancillas are employed, highlighting the effectiveness of the proposed method in exploiting additional ancilla resources.

{ \setlength{\tabcolsep}{6pt}
\begin{table*}[h!tbp]
  \caption{\small{Results for the dynamic realization of the Toffoli operation, comparing the method of \cite{PhysRevA.93.022311} with the proposed clean-ancilla–based dynamic decomposition.}}
  \label{tab1}
   \begin{center}
     \begin{tabular}{r||rrr|rrr|rrr||rrr|rrr|rrr} \toprule 
   & \multicolumn{9}{c||}{n-2 Clean Ancilla} & \multicolumn{9}{c}{$\ceil*{\tfrac{n-2}{2}}$ Clean Ancilla } \\ \cmidrule(lr){2-10}\cmidrule(lr){11-19} 
 & \multicolumn{3}{c|}{Static~\cite{Niel:2000}} & \multicolumn{3}{c|}{Prop. Dyn. W.$^{*}$} & \multicolumn{3}{c||}{Impr.(\%)} & \multicolumn{3}{c|}{Static~\cite{PhysRevA.93.022311}} & \multicolumn{3}{c|}{Prop. Dyn. W.$^{*}$} & \multicolumn{3}{c}{Impr.(\%)}\\ \cmidrule(lr){2-10}\cmidrule(lr){11-19}
$n$& $CX$ & $T_c$ & $T_d$ & $CX$ & $T_c$ & $T_d$ & $CX$ & $T_c$ & $T_d$ & $CX$ & $T_c$ & $T_d$ & $CX$ & $T_c$ & $T_d$ & $CX$ & $T_c$ & $T_d$  \\ \midrule
3 & 12 & 15 & 11 & 10 & 11 & 7 & 16.67 & 26.67 & 36.36 & - & - & - & - & - & - &  &  & \\
4 & 18 & 23 & 11 & 14 & 15 & 7 & 22.22 & 34.78 & 36.36 & 18 & 23 & 19 & 16 & 19 & 15 & 11.11 & 17.39 & 21.05 \\
5 & 24 & 31 & 19 & 18 & 19 & 11 & 25.00 & 38.71 & 42.11 & 24 & 31 & 19 & 20 & 23 & 15 & 16.67 & 25.81 & 21.05 \\
6 & 30 & 39 & 19 & 22 & 23 & 11 & 26.67 & 41.03 & 42.11 & 30 & 39 & 19 & 26 & 31 & 15 & 13.33 & 20.51 & 21.05 \\
7 & 36 & 47 & 27 & 26 & 27 & 15 & 27.78 & 42.55 & 44.44 & 36 & 47 & 27 & 30 & 35 & 21 & 16.67 & 25.53 & 22.22 \\ 
8 & 42 & 55 & 27 & 30 & 31 & 15 & 28.57 & 43.64 & 44.44 & 42 & 55 & 35 & 36 & 43 & 27 & 14.29 & 21.82 & 22.86 \\
9 & 48 & 63 & 35 & 34 & 35 & 19 & 29.17 & 44.44 & 45.71 & 48 & 63 & 35 & 40 & 47 & 27 & 16.67 & 25.40 & 22.86 \\
10 & 54 & 71 & 35 & 38 & 39 & 19 & 29.63 & 45.07 & 45.71 & 54 & 71 & 35 & 46 & 55 & 27 & 14.81 & 22.54 & 22.86 \\
11 & 60 & 79 & 43 & 42 & 43 & 23 & 30.00 & 45.57 & 46.51 & 60 & 79 & 43 & 50 & 59 & 33 & 16.67 & 25.32 & 23.26 \\
12 & 66 & 87 & 43 & 46 & 47 & 23 & 30.30 & 45.98 & 46.51 & 66 & 87 & 51 & 56 & 67 & 39 & 15.15 & 22.99 & 23.53 \\
13 & 72 & 95 & 51 & 50 & 51 & 27 & 30.56 & 46.32 & 47.06 & 72 & 95 & 51 & 60 & 71 & 39 & 16.67 & 25.26 & 23.53 \\
14 & 78 & 103 & 51 & 54 & 55 & 27 & 30.77 & 46.60 & 47.06 & 78 & 103 & 51 & 66 & 79 & 39 & 15.38 & 23.30 & 23.53 \\
15 & 84 & 111 & 59 & 58 & 59 & 31 & 30.95 & 46.85 & 47.46 & 84 & 111 & 59 & 70 & 83 & 45 & 16.67 & 25.23 & 23.73 \\
16 & 90 & 119 & 59 & 62 & 63 & 31 & 31.11 & 47.06 & 47.46 & 90 & 119 & 67 & 76 & 91 & 51 & 15.56 & 23.53 & 23.88 
 \\\bottomrule
\multicolumn{19}{l}{Prop. Dyn. W.$^{*}\xrightarrow{}$ Cost of proposed dynamic realization considering worst case gate overhead.}\\
\multicolumn{19}{l}{$n\xrightarrow{} \mathrm{C}^{n}X$ gate; $\mathrm{C}X \xrightarrow{}\mathrm{C}X$-Count; $T_c\xrightarrow{}T$-Count; $T_d \xrightarrow{}T$-Depth. }\\\bottomrule
      \end{tabular}
      
      
    \end{center}
\end{table*}
}

\section{Conclusion}
In this work, we introduced a dynamic and resource-efficient framework for realizing multi-controlled Toffoli gates in both near-term and fault-tolerant quantum architectures. 
By combining mid-circuit measurement and optimized Clifford+T constructions, the proposed approach systematically reduces entangling-gate count, T-count, and T-depth compared with conventional static decompositions, while preserving fault-tolerance guarantees.

Our analysis and experimental evaluation demonstrate that relative-phase primitives and measurement-adaptive correction strategies enable the construction of high-order Toffoli gates with substantially reduced gate overhead and improved depth scaling. In particular, the proposed dynamic decompositions consistently outperform existing clean-ancilla-based methods under both limited- and abundant-ancilla regimes, achieving lower overall resource overhead even in worst-case execution scenarios.

These results establish dynamic circuit techniques as a practical and scalable approach for implementing reversible logic and arithmetic subroutines that are central to many quantum algorithms. 
By mitigating the dominant non-Clifford and entangling-gate costs, our framework advances the feasibility of large-scale oracle and arithmetic constructions on realistic quantum hardware.

\section{Acknowledgments}
\thanks{This work was partly funded by the Federal Ministry of Research, Technology and Space (BMFTR; formerly BMBF) through the EASEPROFIT project (grant no. 16KIS2127) and by the German Research Foundation (DFG) through the CONAD-QC project (grant no. 559888852). The research is conducted within the scope of the DFG Priority Programme 2514 (SPP 2514).}

\bibliographystyle{IEEEtran}
\bibliography{quantum.bib}
\end{document}